\documentclass{article}

\PassOptionsToPackage{numbers, sort&compress}{natbib}
\usepackage[preprint]{neurips_data_2022}

\usepackage[utf8]{inputenc} 
\usepackage[T1]{fontenc}    %
\usepackage{microtype}
\usepackage{hyperref}
\usepackage{url}
\usepackage{booktabs}
\usepackage{amsmath,amssymb,amsfonts}
\usepackage{algorithmic}
\usepackage{graphicx}
\usepackage{textcomp}
\usepackage{multirow}
\usepackage{multicol}
\usepackage{makecell}
\usepackage{graphicx}
\usepackage{subfigure}
\usepackage{latexsym}
\usepackage{enumitem}
\usepackage{subcaption}
\usepackage{wrapfig}
\mathchardef\mhyphen="2D

\title{{\Large RALL-E: Robust Codec Language Modeling with Chain-of-Thought Prompting for Text-to-Speech Synthesis}}

\author{
\begin{tabular}{c}
Detai Xin$^{1,2}$, Xu Tan$^{1}$, Kai Shen$^{1,3}$, Zeqian Ju$^{1,4}$, Dongchao Yang$^{5}$, Yuancheng Wang$^{6}$\\
Shinnosuke Takamichi$^{2}$, Hiroshi Saruwatari$^{2}$, Shujie Liu$^{1}$, Jinyu Li$^{1}$, Sheng Zhao$^{1}$
\end{tabular}\\
$^{1}$Microsoft, $^{2}$The University of Tokyo\\
$^{3}$Zhejiang University, $^{4}$University of Science and Technology of China\\
$^{5}$The Chinese University of Hong Kong, $^{6}$The Chinese University of Hong Kong, Shenzhen\\
\href{mailto:detai\_xin@ipc.i.u-tokyo.ac.jp}{\texttt{detai\_xin@ipc.i.u-tokyo.ac.jp}} \quad\quad \href{mailto:xuta@microsoft.com}{\texttt{xuta@microsoft.com}}
}

\begin{document}

\maketitle

\vspace{-5mm}
\begin{abstract}
\vspace{-3mm}
We present RALL-E, a robust language modeling method for text-to-speech (TTS) synthesis.
While previous work based on large language models (LLMs) shows impressive performance on zero-shot TTS, such methods often suffer from poor robustness, such as unstable prosody (weird pitch and rhythm/duration) and a high word error rate (WER), due to the autoregressive prediction style of language models.
The core idea behind RALL-E is chain-of-thought (CoT) prompting, which decomposes the task into simpler steps to enhance the robustness of LLM-based TTS.
To accomplish this idea, RALL-E first predicts prosody features (pitch and duration) of the input text and uses them as intermediate conditions to predict speech tokens in a CoT style.
Second, RALL-E utilizes the predicted duration prompt to guide the computing of self-attention weights in Transformer to enforce the model to focus on the corresponding phonemes and prosody features when predicting speech tokens.
Results of comprehensive objective and subjective evaluations demonstrate that, compared to a powerful baseline method VALL-E, RALL-E significantly improves the WER of zero-shot TTS from $5.6\%$ (without reranking) and $1.7\%$ (with reranking) to $2.5\%$ and $1.0\%$, respectively.
Furthermore, we demonstrate that RALL-E correctly synthesizes sentences that are hard for VALL-E and reduces the error rate from $68\%$ to $4\%$.
\end{abstract}

\vspace{-7mm}
\section{Introduction}
\vspace{-3mm}
Large language models (LLMs) have demonstrated great progress in natural language generation~\citep{radford2019gpt2, brown2020gpt3}.
With a sufficient model size LLMs emerge powerful in-context learning abilities that can handle unseen tasks with a text instruction (usually called prompt) in a zero-shot or few-shot manner~\citep{wei2022emergent}.
Moreover, the simple yet effective next-token prediction task of LLMs makes it easy to apply LLMs on other fields, such as vision~\citep{dehghani2023svit} and speech synthesis~\citep{wang2023valle}, as long as the data can be converted to discrete speech tokens.
This work focuses on the language modeling of text-to-speech (TTS) synthesis.
Recent work~\citep{wang2023valle, kharitonov2023speartts} have shown that TTS can be modeled by a decoder-only language model by using a neural codec~\citep{zeghidour2021soundstream, defossez2022encodec} to convert continuous waveforms into discrete tokens.
These methods, typically leverage tens of thousands of hours of speech data, emerge in-context learning ability that can clone a speaker's voice by providing a short audio prompt to the language model, and thus have impressive performance on zero-shot TTS.

\begin{table}[h]
\vspace{-3mm}
    \centering
    \caption{Performance of RALL-E and the baseline method VALL-E~\citep{wang2023valle} on $50$ particularly hard sentences obtained from~\citet{ren2019fastspeech}. The result of NaturalSpeech 2 is from \citet{shen2023ns2}.}
    \footnotesize
    \begin{tabular}{l|cccc|c}
    \toprule
     Model & Mispronunciation & Omission & Repetition & Hallucination & Error rate \\
     \midrule
     NaturalSpeech 2 & $0$ & $0$ & $0$ & $0$ & $0\%$\\
     \midrule
    VALL-E & $10$ & $19$ & $8$ & $7$ & $68\%$ \\
    RALL-E & $2$ & $0$ & $0$ & $0$ & $4\%$\\
    \bottomrule
    \end{tabular}
    \label{tab:hard50}
\end{table}

However, due to the sequential generation property of language models, such codec language models suffer from poor robustness.
Although the autoregressive (AR) prediction style of language models enables the model to generate speech with diverse prosody patterns, they can also cause bad cases with unnatural prosody.
Moreover, since there is no strict alignment between text and speech, the models can omit or repeat words in the input text.
This is quite different from TTS methods based on non-autoregressive (NAR) generative models~\citep{shen2023ns2,le2024voicebox, ju2024ns3}, which predicts all tokens at the same time, thus have high robustness but relatively low diversity.
As suggested by previous work~\citep{yang2023uniaudio, ju2024ns3}, LLM-based TTS have a higher word error rate (WER) than NAR TTS even if they have similar performance on other metrics.
To alleviate this problem, a simple but effective method is to sample the same input text multiple times and select the best one~\citep{kharitonov2023speartts, yang2023uniaudio}.
However, such a reranking method further increases the inference time.

\begin{figure}[t]
    \centering
    \includegraphics[width=\columnwidth]{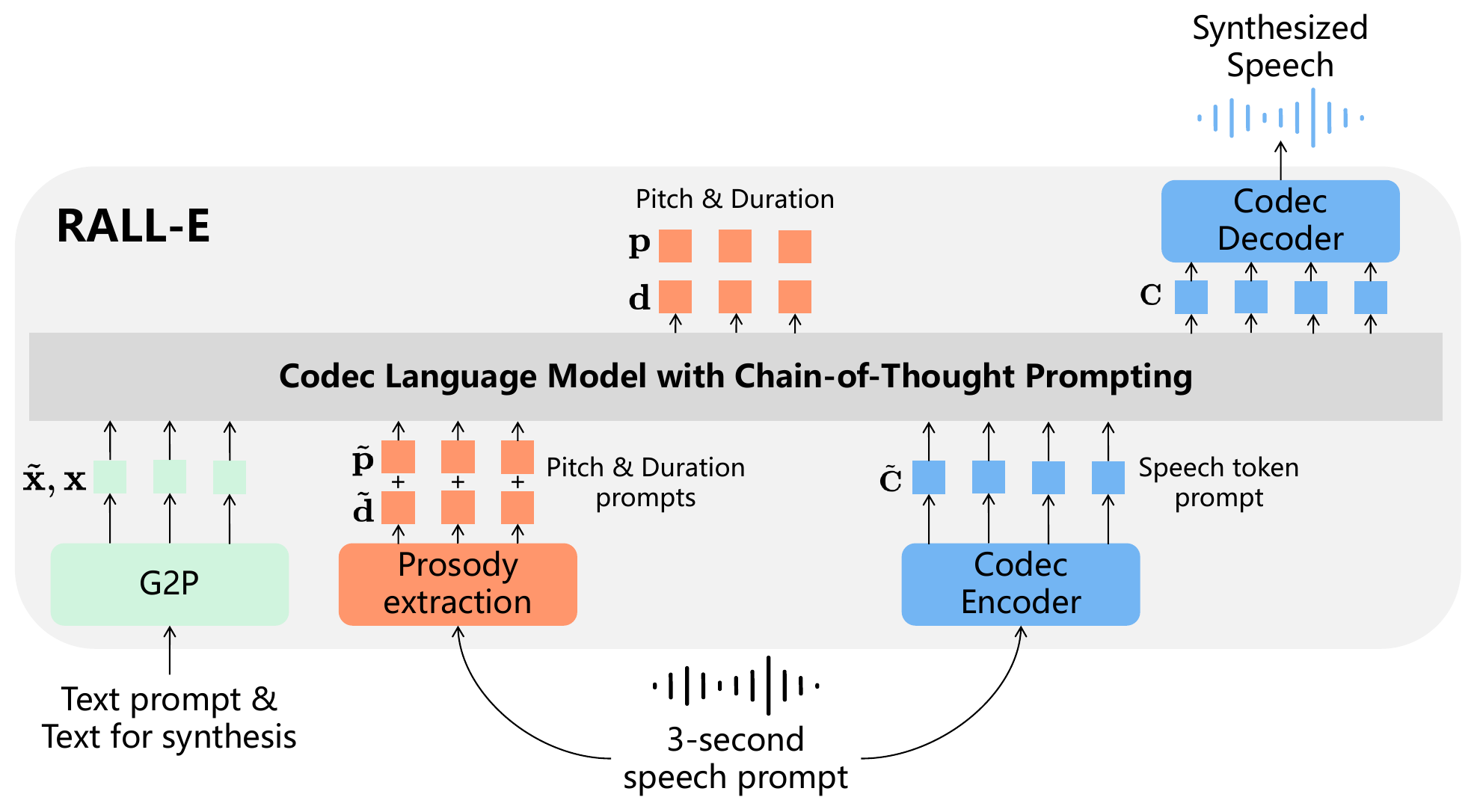}
    \caption{Overview of RALL-E with CoT prompting. Symbols are all defined in Section~\ref{subsection:valle}. The proposed CoT prompting of prosody tokens and duration-guided masking are introduced in Section~\ref{subsection:prosody_prompt} and \ref{subsection:masking}, respectively.}
    \label{figure:overview}
\end{figure}

In this paper, we present RALL-E (the abbreviation of robust VALL-E), a method to improve the robustness of LLM-based TTS.
The core idea of RALL-E is inspired from the chain-of-thought (CoT) prompting~\citep{wei2022chain}.
In CoT prompting, the LLM is instructed to generate an intermediate result that is used as a condition for the prediction of the final result.
The CoT prompting breaks a complex task into several simpler steps, so that can improve the robustness of LLMs, especially on hard tasks like arithmetic~\citep{wei2022chain}.
To adapt CoT prompting to LLM-based TTS, RALL-E predicts prosody tokens (pitch and duration) before predicting speech tokens to stabilize the prosody.
Given an input sentence, RALL-E first predicts phoneme-level pitch and duration of the input, then predicts speech tokens conditioning on both the input phonemes and the predicted prosody tokens.
Furthermore, RALL-E utilizes the predicted duration to mask irrelevant phonemes and prosody tokens when computing self-attention weights, so that the codec language model is enforced to concentrate on tokens around the phoneme and prosody token the speech token corresponds to.
We use VALL-E~\citep{wang2023valle}, a recent powerful LLM-based TTS method, as the baseline, and conduct experiments on a large dataset with $44$K hours speech data.
Results of comprehensive objective and subjective evaluations demonstrate that RALL significantly improves the robustness of LLM-based TTS by reducing the WER on the LibriSpeech~\citep{panayotov2015librispeech} test-clean set from $5.6\%$ (w/o reranking) and $1.7\%$ (with reranking) to $2.5\%$ and $1.0\%$, respectively.
Furthermore, we evaluate RALL-E on $50$ particularly hard sentences.
As demonstrated in Table~\ref{tab:hard50}, compared to VALL-E, RALL-E significantly reduces the error rate from $68\%$ to $4\%$ by eliminating almost all types of error, which demonstrates the superior robustness of RALL-E (see Section~\ref{subsection:hard_sentence} for more details).
The contributions of this work are summarized as follows:
\begin{itemize}[leftmargin=*] 
    \item We present RALL-E, a robust codec language modeling method with chain-of-thought prompting for TTS. RALL-E improves the robustness of LLM-based TTS by (1) incorporating prosody tokens as chain-of-thought prompting to stabilize the generation of speech tokens and (2) using duration-guided masking to enhance the alignment between phoneme and speech tokens.
    \item We conduct comprehensive objective and subjective evaluations. Experimental results demonstrate that RALL-E obtains significantly better robustness than the baseline VALL-E and two previous works.
    \item We further evaluate RALL-E on sentences that are particularly hard to synthesize for LLM-based TTS. The results demonstrate that RALL-E correctly synthesizes hard sentences and reduces the error rate from $68\%$ to $4\%$ compared to VALL-E, which closely approaches the performance of non-autoregressive TTS.
\end{itemize}
Audio samples can be found at \url{https://ralle-demo.github.io/RALL-E}
\section{Related work}
\paragraph{LLM-based TTS}
Inspired by the success of LLMs~\citep{radford2019gpt2, brown2020gpt3}, several recent works adopt language models to model TTS~\citep{wang2023valle,yang2023uniaudio, kharitonov2023speartts} and begin to use decoder-only architecture based on transformer~\citep{vaswani2017attention}.
In such models, text and speech tokens are concatenated together and fed to a single transformer.
The whole model is trained on a next-token prediction task like a language model.
The LLM-based TTS systems are typically trained on tens of thousands of hours of speech data and have hundreds of millions of parameters, hence can leverage the emergent abilities of LLMs like in-context learning~\citep{wei2022emergent} to enable zero-shot TTS~\citep{wang2023valle}.
Besides, recent works~\citep{rubenstein2023audiopalm, wang2023viola, yang2023uniaudio} have shown the decoder-only architecture can be used to learn multiple tasks, as the input and output are processed jointly by a language model, and the model can be signaled to generate results for different tasks by inputting pre-defined special tokens. 
RALL-E focuses on the robustness problem of LLM-based TTS.

\paragraph{Robust autoregressive TTS}
The robustness of AR TTS is a popular topic in the literature.
For encoder-decoder AR TTS, several previous works enforce the attention weights to be monotonic~\citep{zhang2018forward, he2019robust, chen2020multispeech} that can effectively improve the robustness.
In addition, \citet{shen2020robusttacotron} proposed a non-attentive Tacotron, in which the attention module was replaced by a duration predictor to determine the alignment path before decoding.
For decoder-only TTS, a key difference is that the attention weights are computed on text and context at the same time, hence the whole attention weights should not be monotonic.
\citet{song2024ellav} proposed ELLA-V that interleaves the speech tokens with phonemes by inserting a phoneme token and a special \texttt{EndOfPhone} (EOP) token at the beginning and end of the speech tokens corresponding to the phoneme, respectively.
While the inserted phoneme and the EOP token indicate the duration of each phoneme, such an implicit way entangles the prediction of speech tokens and duration together.
RALL-E disentangles the predictions of duration and speech tokens by predicting the duration of all phonemes before the speech tokens, hence has higher controllability over the generation process. 
\citet{du2024vallt} proposed VALL-T that uses an unsupervised transducer loss~\citep{graves2012transducer} to implicitly model the duration of phonemes.
Compared to RALL-E, although VALL-T doesn't rely on external alignment tools during training, its training time is considerably decelerated since the transducer loss requires the model to perform a forward process for every phoneme.
Besides, like ELLA-V, VALL-T also entangles the predictions of duration and speech tokens, thus has weaker controllability than RALL-E.
\section{RALL-E}
The overview of RALL-E is illustrated in Figure~\ref{figure:overview}.
The core idea of RALL-E is CoT prompting that generates intermediate results to assist and stabilize the generation of speech tokens and improve the robustness of LLM-based TTS.
To accomplish this idea, we first propose to predict two kinds of phoneme-level prosody tokens: pitch and duration before predicting the speech tokens.
The distributions of the prosody tokens are modeled together with speech tokens by a single Transformer so that they can influence the duration and pitch of the predicted speech tokens.
To further utilize the predicted duration to guide the generation and improve the robustness, we propose duration-guided masking to enhance the alignment between speech tokens, phonemes, and prosody tokens learned by the language model.
At each decoding step of the speech tokens, RALL-E masks phonemes and prosody tokens that are irrelevant to the synthesis of the current speech token based on the duration information.

In the following sections, we first briefly introduce VALL-E since we apply the proposed method to it in the experiments.
We then formulate and introduce RALL-E in detail.
It should be stressed that, though we use VALL-E to implement RALL-E, the proposed method can be applied in any decoder-only AR TTS model.

\subsection{Preliminary: VALL-E}
\label{subsection:valle}
We inherit most symbols from the original paper of VALL-E~\citep{wang2023valle} for ease of reading.
Readers are recommended to refer to the original paper for more details.

Generally, VALL-E is a decoder-only LLM-based TTS system that uses two Transformers~\citep{vaswani2017attention} to predict speech tokens from the text.
The speech tokens here are extracted from EnCodec~\citep{defossez2022encodec}, a neural audio codec based on residual vector quantization (RVQ)~\citep{zeghidour2021soundstream} that can convert continuous speech signal into discrete tokens.
After predicting the discrete tokens, the waveforms can be reconstructed by feeding the tokens into the decoder of EnCodec.
An RVQ typically contains $N$ quantization layers ($N=8$ in VALL-E), hence at each time step the encoded speech has $N$ tokens.
Formally, given speech $\mathbf{y}$ and its transcription $\mathbf{x}$, the discrete speech token matrix $\mathbf{C}$ encoded by the codec has a shape of $T \times N$, where $T$ is the total time step.
In addition to $\mathbf{x}$, to clone a speaker's voice and utilize the in-context learning ability of LLMs, VALL-E receives a short prompt $\tilde{\mathbf{C}}^{T^{'} \times N}$ as input before predicting $\mathbf{C}$.
Hence, VALL-E models and maximizes the following distribution:
\begin{equation}
\label{eq:valle_overall}
        \mathbb{P} (\mathbf{C}\ |\ \mathbf{x}, \tilde{\mathbf{C}}).
\end{equation}

VALL-E predicts speech tokens hierarchically where the speech tokens of the $1$st layer of RVQ are first predicted by an AR Transformer, and the tokens of the rest layers are predicted by a NAR Transformer.
This is because RVQ uses a residual quantization method, i.e. higher layers encode the information that is not encoded by the lower layers, hence tokens of the $1$st layer contain most information of the waveforms, and the information encoded by the rest layers gradually decreases.
The AR Transformer takes the phoneme sequence $\mathbf{x}$, and speech tokens of the $1$st layer of the prompt $\tilde{\mathbf{c}}_{:, 1}$ as input to predict the target speech tokens of the $1$st layer $\mathbf{c}_{:, 1}$ sequentially, i.e. maximizes the following distribution:
\begin{equation}
\label{eq:valle_ar}
    \mathbb{P} (\mathbf{c}_{:, 1} \ |\ \mathbf{x}, \tilde{\mathbf{c}}_{:, 1}; \theta_{AR}) = \prod_{t=1}^{T} \mathbb{P} (\mathbf{c}_{t, 1} \ |\ \mathbf{x}, \mathbf{c}_{<t, 1}, \tilde{\mathbf{c}}_{:, 1}; \theta_{AR}),
\end{equation}
where $\theta_{AR}$ is the trainable parameters of the AR Transformer.
The NAR Transformer predicts all target speech tokens $\mathbf{c}_{:, j}$ of the $j$th layer at the same time with the phoneme sequence $\mathbf{x}$, the prompt $\tilde{\mathbf{C}}$, and target speech tokens $\mathbf{c}_{:, <j}$ of all layers less than $j$ as the conditions, i.e. maximizes the following distribution:
\begin{equation}
\label{eq:valle_nar}
    \mathbb{P} (\mathbf{c}_{:, 2:N} \ |\ \mathbf{x}, \tilde{\mathbf{C}} ; \theta_{NAR}) = \prod_{j=2}^{N} \mathbb{P} (\mathbf{c}_{:, j} \ |\ \mathbf{x}, \mathbf{c}_{:, <j},\tilde{\mathbf{C}} ; \theta_{NAR}),
\end{equation}
where $\theta_{NAR}$ is the trainable parameters of the NAR Transformer.
By combining Eq.~\ref{eq:valle_ar} and Eq.~\ref{eq:valle_nar} VALL-E breaks Eq.~\ref{eq:valle_overall} into the following form:
\begin{equation}
    \mathbb{P} (\mathbf{C} \ |\ \mathbf{x}, \tilde{\mathbf{C}}) = \mathbb{P} (\mathbf{c}_{:, 1} \ |\ \mathbf{x}, \tilde{\mathbf{c}}_{:, 1}; \theta_{AR}) \mathbb{P} (\mathbf{c}_{:, 2:N} \ |\ \mathbf{x}, \tilde{\mathbf{C}} ; \theta_{NAR}).
\end{equation}
It is noteworthy that in practice the two Transformers have the same architecture but have different attention masks during computation.
Specifically, both the two Transformers use a bidirectional mask for the phoneme sequence $\mathbf{x}$, which means every phoneme $x_{i}$ can attend to all other phonemes $x_{\neq i}$.
However, for the speech tokens, the AR Transformers uses a unidirectional mask so that $\mathbf{c}_{t, 1}$ can only attend to previous tokens $\mathbf{c}_{<t, 1}$, while the NAR Transformer still uses a bidirectional mask.
\subsection{Prosody tokens as chain-of-thought prompts}
\label{subsection:prosody_prompt}
One of the problems of LLM-based TTS is that it directly generates speech from phonemes with no restriction on the prosody, e.g. pitch, duration, etc, which usually results in speech with unstable prosody.
A similar problem is also observed in \citet{wei2022chain} where the authors find LLMs cannot directly answer a complex question like arithmetic and propose CoT prompting to solve this problem.
The idea of CoT prompting is breaking a complex task into several simpler tasks so that LLMs can utilize the intermediate results to reach the final answers.
As shown in \citet{wei2022chain}, by CoT prompting the correct rate of LLMs on complex tasks can be significantly improved.
This motivates us to adapt CoT prompting to LLM-based TTS by generating intermediate prosody tokens before generating speech tokens to alleviate the robustness problem of LLM-based TTS.
To incorporate pitch and duration in the AR Transformer of VALL-E, we first get the alignment between phonemes and speech tokens and extract the pitch value for each speech token.
We then compute phoneme-level pitch value based on the duration and linearly quantize it to $M_{p}$ buckets.
We define a maximal duration value $M_{d}$, and all duration values that exceed $M_{d}$ will be truncated to the maximum.
RALL-E predicts the two prosody tokens before the speech tokens in a CoT style.
Formally, assume $\mathbf{p}, \mathbf{d}$ are the discrete pitch and duration sequences of the target speech tokens $\mathbf{C}$, $\tilde{\mathbf{p}}, \tilde{\mathbf{d}}$ are the ones of the prompt $\tilde{\mathbf{C}}$, we model and maximize the following distribution:
\begin{equation}
\label{eq:prop_pd}
    \mathbb{P} (\mathbf{p}, \mathbf{d} \ |\ \mathbf{x}, \tilde{\mathbf{p}}, \tilde{\mathbf{d}} ; \theta_{AR}) = \prod_{t=1}^{L} \mathbb{P} (p_{t}, d_{t} \ |\ \mathbf{x}, \mathbf{p}_{<t}, \mathbf{d}_{<t}, \tilde{\mathbf{p}}, \tilde{\mathbf{d}} ; \theta_{AR}),
\end{equation}
where $L$ is the length of $\mathbf{x}$.
In practice, the model predicts $p_{t}$ and $d_{t}$ with two separate heads, and their embeddings are summed up and fed to the model for the prediction of the next step.
RALL-E then predicts the speech tokens with $\mathbf{p}$ and $\mathbf{d}$ as a new condition, which makes Eq.~\ref{eq:valle_ar} becomes:
\begin{equation}
\label{eq:prop_ar}
        \mathbb{P} (\mathbf{c}_{:, 1} \ |\ \mathbf{x}, \tilde{\mathbf{c}}_{:, 1}, \mathbf{p}, \tilde{\mathbf{p}}, \mathbf{d}, \tilde{\mathbf{d}}; \theta_{AR}) = \prod_{t=1}^{T} \mathbb{P} (\mathbf{c}_{t, 1} \ |\ \mathbf{x}, \mathbf{c}_{<t, 1}, \tilde{\mathbf{c}}_{:, 1}, \mathbf{p}, \tilde{\mathbf{p}}, \mathbf{d}, \tilde{\mathbf{d}} ; \theta_{AR}).
\end{equation}
The above two equations can be jointly optimized by the AR Transformer.
Although the proposed method adds additional $L$ decoding steps, since $L \ll T$, it intuitively has little influence on the efficiency.

For the NAR Transformer, we simply sum the embeddings of the phoneme, pitch, and duration together as the input.
This makes Eq.~\ref{eq:valle_nar} becomes:
\begin{equation}
\label{eq:prop_nar}
    \mathbb{P} (\mathbf{c}_{:, 2:N} \ |\ \mathbf{x}, \tilde{\mathbf{C}}, \mathbf{p}, \tilde{\mathbf{p}}, \mathbf{d}, \tilde{\mathbf{d}} ; \theta_{NAR}) = \prod_{j=2}^{N} \mathbb{P} (\mathbf{c}_{:, j} \ |\ \mathbf{x}, \mathbf{c}_{:, <j},\tilde{\mathbf{C}}, \mathbf{p}, \tilde{\mathbf{p}}, \mathbf{d}, \tilde{\mathbf{d}} ; \theta_{NAR}).
\end{equation}

\begin{figure}[t]
    \centering
    \includegraphics[width=0.9\columnwidth]{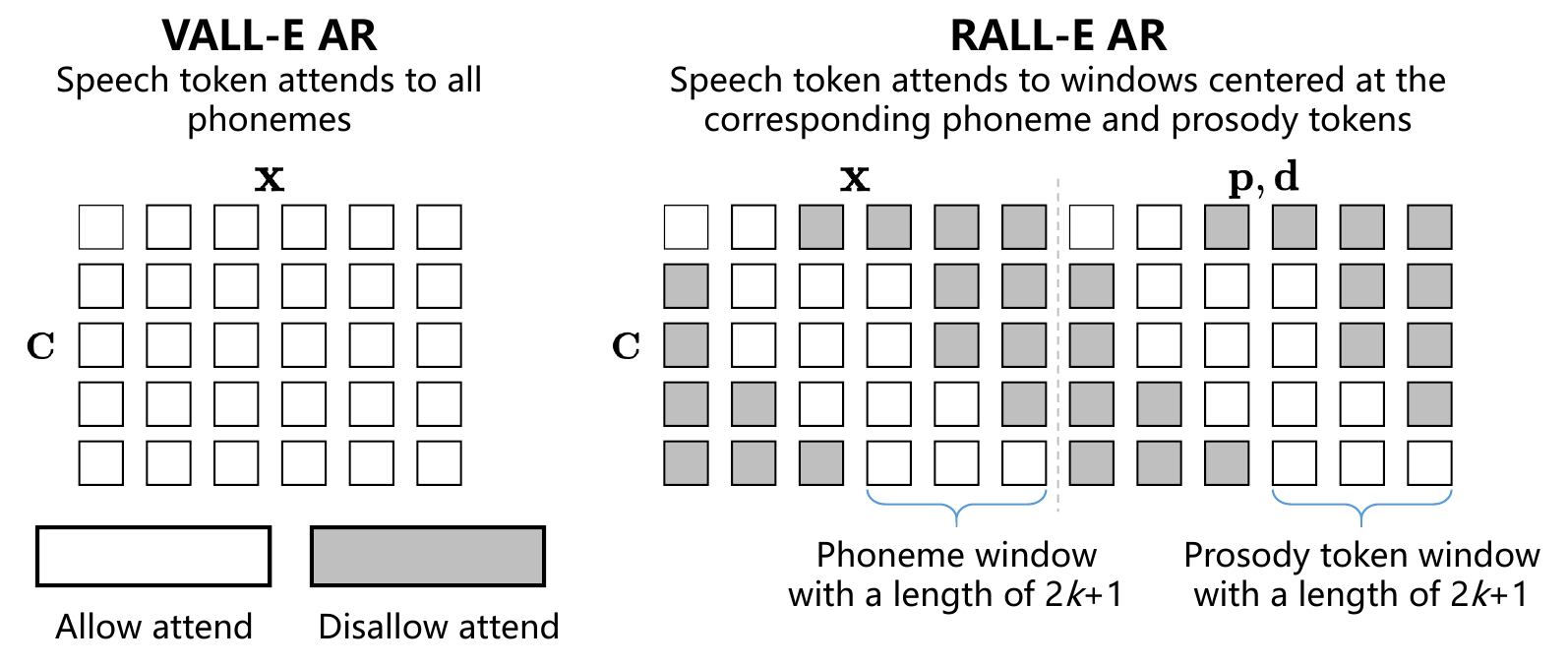}
    \caption{A comparison between how speech token attends to phonemes in the AR Transformer of VALL-E and RALL-E. Here $k$ is set to $1$.}
    \label{figure:masking}
\end{figure}

\subsection{Enhancing alignment with duration-guided masking}
\label{subsection:masking}
As the left side of Figure~\ref{figure:masking} illustrates, since the speech token attends to all phonemes in the AR Transformer of VALL-E, the alignment between the phonemes and the speech tokens is implicitly modeled by the self-attention of VALL-E.
This can be imprecise and causes errors like word omissions or hallucinations.
Though RALL-E introduces prosody CoT prompting to guide and stabilize the generation, we still find the model can fail to align in the experiments.
We thus propose duration-guided masking to fully utilize the intermediate duration results and boost the robustness.

As the right side of Figure~\ref{figure:masking} illustrates, in the proposed duration-guided masking, the speech token is restricted to only attend on a phoneme (prosody token) window centered at the phoneme (prosody token) it corresponds to.
We define the window size as $k$, thus each speech token can attend on $2k+1$ phonemes and $2k+1$ prosody tokens. 
All phonemes and prosody tokens at other positions will be masked out, hence their attention weights are always zero.
When $k=0$ the speech token strictly attends to the phoneme it corresponds to.
If the alignment is perfect this should be enough.
However, in the experiments, we found that the alignment results obtained by our alignment tool usually have errors.
We thus loosen the restriction by also allowing the speech token to attend at the near phonemes of the corresponding phoneme.
Another reason for this design is that the pronunciation of a phoneme is usually dependent on near phonemes.
As one will see in Section~\ref{subsection:ablation} and Appendix~\ref{appendix:window_size}, the experimental results verify the effectiveness of this design.
For the NAR Transformer, we obtained almost no gain when applying the proposed masking strategy to it in our preliminary experiments.
Thus we only apply the masking strategy on the AR Transformer.

The general inference procedure follows VALL-E~\citep{wang2023valle} with two differences.
First, before sampling the speech tokens $\mathbf{c}_{:, 1}$ the prosody tokens $\mathbf{p}$ and $\mathbf{d}$ are sampled conditioning on the phoneme sequence $\mathbf{x}$ and acoustic prompt $\tilde{\mathbf{p}}, \tilde{\mathbf{d}}$.
Second, although normal language models depend on a special token \texttt{<eos>} as the stop condition, since we know the total duration $D=\sum_{t=1}^{L}d_{t}$, we propose a duration-guided inference method that forces the inference to stop at the $D$-th step.
This method ensures no phoneme is omitted or repeated as it continues the inference if the \texttt{<eos>} token is predicted before the $D$-th step and stops at the right step as guided by the predicted duration.
\section{Experiments}
\label{section:experiments}
\subsection{Setup}
\paragraph{Data}
We use the English subset of the multilingual LibriSpeech (MLS)~\citep{pratap20mls} corpus containing about $44$K hours of speech data from $5490$ speakers as the training data.
The test-clean set of the LibriSpeech corpus~\cite{panayotov2015librispeech} is used as the test set.
Following \citet{wang2023valle}, we only select utterances with a length of $4$-second to $10$-second, which results in $1205$ utterances.
For each utterance in the test set, we randomly select another one of the same speaker and use the first $3$s of it as the prompt.
All speech data are in $16$~kHz.
The transcriptions are converted into phonemes by a grapheme-to-phoneme tool~\citep{sun19graphemetophoneme}.
The frame-level pitch value is extracted by the WORLD vocoder~\citep{morise2016world}.
We get the alignments between the phoneme sequences and the speech tokens with our internal alignment tool.
The maximal duration value $M_{d}$ is set to $32$.
The phoneme-level pitch value is then computed based on the alignments.
The number of buckets $M_{p}$ for pitch quantization is set to $256$.

\paragraph{Model configuration}
We use SoundStream~\citep{zeghidour2021soundstream} as the speech codec to get speech tokens and decode waveforms from the tokens.
The architecture follows the original work and the number of quantization layer $N$ of RVQ is set to $16$.
The codec language model follows VALL-E~\citep{wang2023valle}, where both the AR and NAR models are a $12$-layer Transformer~\citep{vaswani2017attention} with $1024$-dimensional token embeddings, sinusoidal positional embeddings, $4096$-dimensional feed-forward layers, and a dropout rate of $0.1$.
The window size $k$ is set to $1$ without explicit statements (see Appendix~\ref{appendix:window_size} for a detailed explanation of how we choose this value).

\paragraph{Training and inference}
The SoundStream codec is trained on $8$ NVIDIA V100 GPUs with a batch size of $200$ per GPU.
We use Adam~\citep{kingma2014adam} as the optimizer with a learning rate of $2e\mhyphen4$.
The model converges in about $440$K steps.
The AR and NAR Transformers are trained separately on $16$ AMD MI200 GPUs with a batch size of $7000$ speech tokens per GPU.
Adam is again used as the optimizer.
The scheduled inverse square root learning rate is used with $30$K warm-up steps and a peak learning rate of $5e\mhyphen4$.
Both of the two Transformers converge in about $500$K steps.

We adopt nucleus sampling~\citep{holtzman2019nucleus} as the sampling method of the AR Transformer.
For the predicted probability distribution, nucleus sampling first selects a token set with the highest probability whose accumulative probability exceeds a hyperparameter $\rho$, and randomly samples from the set.
Note that, $\rho$ can be different for the sampling of pitch, duration, and speech tokens, which results in $3$ hyperparameters: $\rho_{p}$, $\rho_{d}$, $\rho_{c}$ for pitch, duration, speech tokens, respectively.
Without explicit statements we set $\rho_{p} = \rho_{d} = \rho_{c} = 0.9$ in the following experiments.
We select the token with the highest probability in the NAR Transformer without sampling.

\paragraph{Baseline methods}
We use VALL-E~\citep{wang2023valle} as the baseline method.
We implement VALL-E and train it on our training set.
Besides, two previous works: VALL-T~\citep{du2024vallt} and ELLA-V~\citep{song2024ellav} are also compared with RALL-E.
VALL-T is trained on the LibriTTS~\citep{zen2019libritts} corpus with $520$ hours of speech data, and we get $500$ synthesized samples selected from the test-clean set of LibriTTS from the authors.
ELLA-V~\citep{song2024ellav} is trained on the training set of the LibriSpeech~\cite{panayotov2015librispeech} corpus with $960$ hours speech data.
We request the authors to run ELLA-V on our test set and get $912$ samples.
We don't use the result of the original paper~\citep{song2024ellav} since the continual generation method adopted by ELLA-V can result in a significantly better WER than non-continual generation, as suggested by \citet{wang2023valle}.

\begin{table}[t]
    \centering
    \caption{Main results of RALL-E on the LibriSpeech test set with $1205$ utterances. \textbf{Bold} indicates the best score. The WER shown in the parenthesis is obtained from the HuBERT model used in the original VALL-E paper~\citep{wang2023valle}. $^\dagger$ means the results of VALL-E trained on LibriLight. $^\ddagger$ means the results of VALL-E trained on the English subset of MLS. We get the results of VALL-T ($500$ samples)~\citep{du2024vallt} and ELLA-V ($912$ samples)~\citep{song2024ellav} from the authors. RALL-E ($912$) means the results are computed on the same test set with ELLA-V including $912$ samples.}
    \footnotesize
    \begin{tabular}{l|cccc|ccc}
    \toprule
     & WER\% ($\downarrow$)  & WER-R\% ($\downarrow$) & UTMOS ($\uparrow$) & SIM ($\uparrow$) & Sub ($\downarrow$) & Del ($\downarrow$) & Ins ($\downarrow$) \\
    \midrule
    GT       & $1.8$ ($2.1)$ & - & $4.1$  & $0.69$ & $1.4$ & $0.2$ & $0.2$ \\
    VALL-E$^\dagger$  & - ($5.9$) & - & - & $0.58$ & - & - & -\\
    \midrule
    VALL-E$^\ddagger$ & $5.6$ ($6.3)$ & $1.7$ & $3.9$ & $0.49$ & $2.8$ ($3.6$) & $1.5$ ($1.4$) & $1.3$  ($1.3$) \\
    ELLA-V ($912$)  & $2.8$ ($4.1$)  & $0.8$  & $3.7$ & $0.42$ & $2.2$ ($3.4$) & $0.4$ ($0.4$) & $0.2$ ($0.3$) \\
    VALL-T ($500$) & $3.9$ ($5.4$) & - & $\mathbf{4.0}$ & $0.46$ & $2.4$ ($3.6$) & $1.3$ ($1.6$) & $\mathbf{0.2}$ ($0.2$) \\
    \midrule
    RALL-E ($912$) & $2.3$ ($2.6$) & $0.8$ & $4.0$ & $0.49$ & $1.4$ ($2.0$) & $0.6$ ($0.3$) & $0.3$ ($0.3$) \\
    RALL-E  & $\mathbf{2.5}$ ($\mathbf{2.8}$) & $\mathbf{1.0}$ & $\mathbf{4.0}$ & $0.49$ & $\mathbf{1.7}$ ($2.2$) & $0.6$ ($0.3$) & $0.3$ ($0.3$) \\
    \bottomrule
    \end{tabular}
    \label{tab:main_result}
\end{table}
\paragraph{Objective metrics}
We use the following objective metrics:
\begin{itemize}[leftmargin=*] 
    \item \textbf{Word error rate (WER)}. We transcribe the synthesized samples by a large Conformer-based~\citep{gulati20conformer} ASR model\footnote{\url{https://huggingface.co/nvidia/stt_en_conformer_transducer_xlarge}} that is trained on a large collection of speech corpora including LibriSpeech~\citep{panayotov2015librispeech}. The WER is then computed between the recognized and the ground truth (GT) transcriptions. Besides, we also report WERs computed from transcriptions recognized by a  HuBERT\footnote{\url{https://huggingface.co/facebook/hubert-large-ls960-ft}}~\citep{hsu2021hubert} model trained on Libri-Light~\citep{2020librilight} and finetuned on LibriSpeech~\citep{panayotov2015librispeech}. We regard the WERs obtained from the Conformer-based model as the primary scores since it results in better performance than the HuBERT model, though the HuBERT model is used in the original VALL-E paper~\citep{wang2023valle}.
    \item \textbf{Reranked WER (WER-R)}. For each test utterance, we sample it $5$ times and select the one with the best edit distance to the GT transcription to compute WER. This metric can be regarded as a performance upper bound, while the WER represents average performance.
    \item \textbf{Substitution (Sub)}, \textbf{Deletion (Del)}, and \textbf{Insertion (Ins)} computed by the edit distance algorithm. These three metrics are by-products of the computing of WER and can reveal specific error types made by the TTS model. Typically Sub refers to mispronunciation, Del refers to word omission, and Ins refers to word repetition or hallucination.
    \item \textbf{UTMOS}~\citep{saeki2022utmos}, which is a powerful automatic speech quality assessment model that evaluates speech naturalness.
    \item \textbf{Speaker similarity (SIM)} defined as the cosine similarity between the speaker embeddings of the prompt and the synthesized utterance. Following VALL-E~\citep{wang2023valle}, we use the \texttt{
wavlm\_large\_finetune} checkpoint of WavLM-TDNN\footnote{\url{https://github.com/microsoft/UniSpeech/tree/main/downstreams/speaker_verification}}, a speaker verification model based on WavLM~\citep{chen2022wavlm}, to extract speaker embeddings.
\end{itemize}
\paragraph{Subjective metrics}
We use two common subjective metrics: comparative mean opinion score (CMOS) and similarity mean opinion score (SMOS) to evaluate speech naturalness and speaker similarity, respectively.
CMOS shows the performance difference between two systems ranging from $-3$ (the new system is much worse than the old system) to $3$ (the new system is much better than the old system).
SMOS has $5$ scales from $1$ to $5$, and higher values indicate better speaker similarity between the synthesized and GT samples.

\subsection{Main results}
We first evaluate the overall performance of RALL-E on the full LibriSpeech test set with $1205$ utterances.
The results are demonstrated in Table~\ref{tab:main_result}.
It can be observed that RALL-E outperforms all other methods on WER.
The reranked WER (WER-R) of RALL-E is even better than the WER of GT.
Compared to the baseline VALL-E method, RALL-E obtains a $55\%$ relative improvement on WER and a $41\%$ relative improvement on WER-R, showing the superior robustness of the proposed method.
This is also proved by the performance on the three error types, where RALL-E consistently reduces all error types from $2.8$/$1.5$/$1.3$ to $1.7$/$0.6$/$0.3$, respectively.
In addition, the higher UTMOS score of RALL-E than the one of VALL-E indicates RALL-E can synthesize speech with better naturalness, showing the effectiveness of the proposed method on stabilizing speech prosody.
Regarding the speaker similarity, both RALL-E and VALL-E show better performance than the previous two works which is possibly because of the larger training set used by us.
However, the original VALL-E has a significantly higher SIM score ($0.58$) than other methods.
One possible reason is that the speaker similarity of the original VALL-E is computed between the synthesized utterance and the prompt resynthesized by the codec instead of the GT prompt.
We notice that VALL-T has slightly fewer insertion errors ($0.2$) than RALL-E ($0.3$), but we think this is reasonable and doesn't necessarily mean RALL-E is worse than VALL-T since the test set of VALL-T only contains $500$ samples and fewer test samples usually result in a better WER, as suggested by the result of RALL-E ($912$) computed on the $912$ samples used by ELLA-V.

\begin{wraptable}{r}{0.5\textwidth}
    \centering
    \caption{Results of subjective CMOS (v.s. RALL-E) and SMOS tests. \textbf{Bold} indicates the best score.}
    \footnotesize
    \begin{tabular}{l|cc}
    \toprule
     & CMOS & SMOS \\
     \midrule
     GT & -0.02  &  $4.23$\\
     \midrule
     VALL-E & -0.17 &  $3.50$ \\
     RALL-E & ~\textbf{0.00}  & $3.57$ \\
    \bottomrule
    \end{tabular}
    \label{tab:sub_eval}
\end{wraptable}
Next, we conduct subjective tests to evaluate the performance of RALL-E.
We randomly select $20$ samples from the test set for the CMOS tests, and $10$ samples from distinct speakers for the SMOS test.
Two CMOS tests with $6$ workers per test are conducted on two pairs: (GT, RALL-E) and (VALL-E, RALL-E), respectively.
Thus each utterance has $6$ answers.
Similarly, an SMOS test is conducted with $6$ workers.
The results are shown in Table~\ref{tab:sub_eval}.
It can be seen that RALL-E obtains a better CMOS than VALL-E and even slightly better performance than GT utterance, showing the effectiveness of RALL-E on stabilizing prosody by incorporating prosody tokens as CoT prompting.
Regarding SMOS the two methods have similar performance, which is consistent with the result of SIM scores shown in Table~\ref{tab:main_result}.

\subsection{Ablation study}
\label{subsection:ablation}
\begin{table}[t]
    \centering
    \caption{Results of ablation studies. \textbf{Bold} indicates the best score.}
    \footnotesize
    \begin{tabular}{l|cc|ccc}
    \toprule
     & WER\%($\downarrow$) & UTMOS ($\uparrow$) & Sub ($\downarrow$) & Del ($\downarrow$) & Ins ($\downarrow$)   \\
     \midrule
     RALL-E & $\mathbf{2.5}$ & $\mathbf{4.00}$ & $\mathbf{1.7}$ & $\mathbf{0.5}$  & $\mathbf{0.3}$ \\
     \midrule
     w/o pitch & $2.6$ & $3.96$ & $1.8$ & $0.5$ & $0.3$ \\
     w/o window masking & $2.7$ & $3.84$ & $1.8$ & $0.6$     & $0.3$ \\
     w/o duration-guided masking & $3.2$ & $3.88$ & $2.0$ & $0.8$ & $0.5$ \\
     w/o duration CoT prompting & $13.4$ & $3.52$ & $7.8$ & $4.1$ & $1.5$ \\
    \bottomrule
    \end{tabular}
    \label{tab:ablation}
\end{table}
We conduct ablation experiments to study the contributions of each component of RALL-E.
Specifically, we study the following four settings: (1) w/o pitch that removes the pitch prompt; (2) w/o window masking where the window size $k$ is set to $0$; (3) w/o duration-guided masking that uses normal unidirectional autoregressive attention masks; (4) w/o duration CoT prompting that removes duration from the CoT and model it separately.
In the w/o duration CoT prompting setting we use an independent $8$-layer Transformer with $256$-dimensional token embeddings and $8$ self-attention heads to predict duration, the masking strategy is still used based on the duration.

The results are shown in Table~\ref{tab:ablation}.
First, the result of w/o pitch demonstrates that the pitch token helps to reduce mispronunciation.
Second, it can be observed from the result of w/o window masking that the model performance degrades when the window size is set to $0$, which demonstrates the effectiveness of the window masking strategy.
See Appendix~\ref{appendix:window_size} for a comprehensive study on the effects of window size $k$.
Third, w/o duration-guided masking shows consistently worse performance than RALL-E on all metrics, showing the effectiveness of the proposed duration-guided masking strategy.
Finally, w/o duration CoT prompting has the worst performance even though it uses the proposed duration-guided masking.
By manually listening to the samples we find that the predicted duration of the independent Transformer does not influence the synthesized speech, which demonstrates the necessity of incorporating the duration with CoT prompting.
In summary, each component of RALL-E contributes to the robustness improvements, and the CoT prompting is the most important component of RALL-E.

\subsection{Evaluations on hard sentences}
\label{subsection:hard_sentence}
To further evaluate the robustness of RALL-E, we synthesize $50$ particularly hard sentences (see Appendix~\ref{appendix:hard_sentence} for the transcripts of these sentences) with RALL-E and VALL-E.
We manually evaluate the results since the WER computed on these sentences with a lot of numbers and symbols is imprecise.
We decompose the possible errors into four types: mispronunciation, omission, repetition, and hallucination.
Each utterance is synthesized $5$ times and the best one is selected.
We count the frequency of all error types and compute the overall sentence error rate.
Each error type is only counted once in an utterance.
The results are shown in Table~\ref{tab:hard50}, in which we also add a powerful non-autoregressive TTS method NaturalSpeech2~\citep{shen2023ns2} for reference.
It can be seen that RALL-E substantially reduces the error rate from $68\%$ to $4\%$ with only $2$ mispronunciation errors.
This performance approaches the one of non-autoregressive NaturalSpeech2 that has no error, which again demonstrates the effectiveness of RALL-E in improving the robustness of LLM-based TTS.
Specifically, we observe that for extremely short sentences, e.g. a single alphabet ``A'', VALL-E usually synthesizes words that don't exist in the input sentence, causing a severe hallucination problem.
Besides, if a word is repeated many times in a sentence, e.g. ``22222222'', VALL-E easily makes mistakes by omitting or repeating the word.
These problems together demonstrate that LLM-based TTS like VALL-E has low controllability on the duration of the synthesized speech and learns poor alignments between phonemes and speech tokens.
In contrast, RALL-E first introduces prosody tokens as CoT prompting to improve duration controllability and then uses duration-guided masking to enhance the alignment, which effectively alleviates the aforementioned problems made by the language models.
All in all, our RALL-E demonstrates superior robustness in all evaluations.
\section{Conclusions}
This paper presents RALL-E, a robust codec language modeling method with CoT prompting for TTS.
To address the robustness problem of LLM-based TTS, RALL-E (1) incorporates prosody features (pitch and duration) in the LLM as a CoT prompting to assist and stabilize the generation of speech tokens, and (2) proposes duration-guided masking that enforces the model to attend on relevant phonemes (prosody features) corresponding to each speech token.
We conduct comprehensive objective and subjective evaluations and demonstrate that RALL-E can significantly improve the robustness of LLM-based TTS compared to the baseline VALL-E and two previous works.
Furthermore, we show that RALL-E can correctly synthesize sentences that are particularly hard to synthesize for VALL-E with a $4\%$ error rate that even approaches the performance of non-autoregressive TTS.
\begin{ack}
We thank Yakun Song and Chenpeng Du for providing the test samples of ELLA-V and VALL-T, respectively.
\end{ack}

\bibliographystyle{abbrvnat}
\bibliography{colm2024_conference}



\appendix
\section{Window size study}
\label{appendix:window_size}
\begin{wrapfigure}{R}{0.5\linewidth}
    \centering
    \includegraphics[width=0.48\columnwidth]{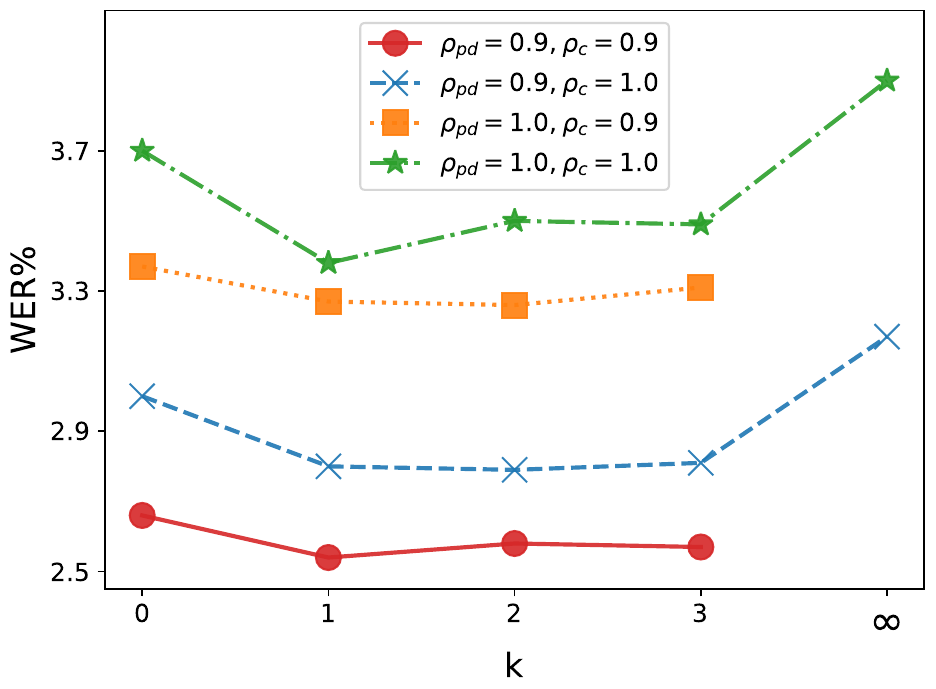}
    \caption{Results of the window size study. For simplicity, we use $\rho_{pd}$ to refer to $\rho_{p}$ and $\rho_{d}$ together. $k=\infty$ means the window covers the whole phoneme sequence, i.e. no phoneme is masked. The model with $k=\infty$ fails to generate speech tokens when $\rho_{c}=0.9$.}
    \label{figure:window_size}
\end{wrapfigure}
We study the window size hyperparameter $k$ used in the proposed duration-guided masking method.
Basically, $2k+1$ is the number of phonemes (prosody features) the model can attend during decoding.
As mentioned in Section~\ref{subsection:masking}, the motivation of the window is to (1) increase context information received by the model during decoding and (2) improve the robustness of the proposed duration-based masking strategy since the extracted duration features can have errors during training and the predicted duration may not strictly correspond to the number of predicted speech tokens for each phoneme during inference.
We suppose $k=0$ will make RALL-E less robust, but large $k$ will also make it difficult to learn the alignment between phonemes and speech tokens.
Thus we study the optimal value of $k$.
We train RALL-E and compute WER on the test set with $k=0,1,2,3,\infty$, in which $k=\infty$ means the window covers the whole phoneme sequence, i.e. no phoneme is masked during decoding.
We hypothesize that diverse sampling will make the robustness problem more obvious, thus we perform nucleus sampling on each model with four settings: (1) $\rho_{p}=\rho_{d}=\rho_{c}=0.9$; (2) $\rho_{p}=\rho_{d}=0.9, \rho_{c}=1.0$; (3) $\rho_{p}=\rho_{d}=1.0, \rho_{c}=0.9$; and (4) $\rho_{p}=\rho_{d}=1.0, \rho_{c}=1.0$.
The results are illustrated in Figure~\ref{figure:window_size}.
First, it can be observed that in every sampling setting the WER can be substantially improved by increasing $k$ from $0$ to $1$, showing the effectiveness of the proposed window masking strategy.
This observation also verifies the hypothesis that the more the sampling becomes diverse, the more the robustness problem becomes obvious.
Second, the performance cannot be further improved by increasing $k$ to values larger than $1$, which verifies another hypothesis that large $k$ makes it difficult to learn the alignment.
When $k=\infty$ the model has to learn the alignment completely by itself, thus resulting in the worst WERs.
Combining all results we conclude that the proposed window masking strategy can effectively improve WERs and the best performance is obtained when $k=1$.

\section{Transcripts of the 50 hard sentences}
\label{appendix:hard_sentence}
We list the $50$ hard sentences used in Section~\ref{subsection:hard_sentence} below:
\begin{enumerate}[leftmargin=*] \itemsep -1mm
\item a
\item b
\item c
\item H
\item I
\item J
\item K
\item L
\item 22222222 hello 22222222
\item S D S D Pass zero - zero Fail - zero to zero - zero - zero Cancelled - fifty nine to three - two - sixty four Total - fifty nine to three - two -
\item S D S D Pass - zero - zero - zero - zero Fail - zero - zero - zero - zero Cancelled - four hundred and sixteen - seventy six -
\item zero - one - one - two Cancelled - zero - zero - zero - zero Total - two hundred and eighty six - nineteen - seven -
\item forty one to five three hundred and eleven Fail - one - one to zero two Cancelled - zero - zero to zero zero Total -
\item zero zero one , MS03 - zero twenty five , MS03 - zero thirty two , MS03 - zero thirty nine ,
\item 1b204928 zero zero zero zero zero zero zero zero zero zero zero zero zero zero one seven ole32 11
\item zero zero zero zero zero zero zero zero two seven nine eight F three forty zero zero zero zero zero six four two eight zero one eight
\item c five eight zero three three nine a zero bf eight FALSE zero zero zero bba3add2 - c229 - 4cdb -
\item Calendaring agent failed with error code 0x80070005 while saving appointment .
\item Exit process - break ld - Load module - output ud - Unload module - ignore ser - System error - ignore ibp - Initial breakpoint -
\item Common DB connectors include the DB - nine , DB - fifteen , DB - nineteen , DB - twenty five , DB - thirty seven , and DB - fifty connectors .
\item To deliver interfaces that are significantly better suited to create and process RFC eight twenty one , RFC eight twenty two , RFC nine seventy seven , and MIME content .
\item int1 , int2 , int3 , int4 , int5 , int6 , int7 , int8 , int9 ,
\item seven \_ ctl00 ctl04 ctl01 ctl00 ctl00
\item Http0XX , Http1XX , Http2XX , Http3XX ,
\item config file must contain A , B , C , D , E , F , and G .
\item mondo - debug mondo - ship motif - debug motif - ship sts - debug sts - ship Comparing local files to checkpoint files ...
\item Rusbvts . dll Dsaccessbvts . dll Exchmembvt . dll Draino . dll Im trying to deploy a new topology , and I keep getting this error .
\item You can call me directly at four two five seven zero three seven three four four or my cell four two five four four four seven four seven four or send me a meeting request with all the appropriate information .
\item Failed zero point zero zero percent < one zero zero one zero zero zero zero Internal . Exchange . ContentFilter . BVT ContentFilter . BVT\_log . xml Error ! Filename not specified .
\item C colon backslash o one two f c p a r t y backslash d e v one two backslash oasys backslash legacy backslash web backslash HELP
\item src backslash mapi backslash t n e f d e c dot c dot o l d backslash backslash m o z a r t f one backslash e x five
\item copy backslash backslash j o h n f a n four backslash scratch backslash M i c r o s o f t dot S h a r e P o i n t dot
\item Take a look at h t t p colon slash slash w w w dot granite dot a b dot c a slash access slash email dot
\item backslash bin backslash premium backslash forms backslash r e g i o n a l o p t i o n s dot a s p x dot c s Raj , DJ ,
\item Anuraag backslash backslash r a d u r five backslash d e b u g dot one eight zero nine underscore P R two h dot s t s contains
\item p l a t f o r m right bracket backslash left bracket f l a v o r right bracket backslash s e t u p dot e x e
\item backslash x eight six backslash Ship backslash zero backslash A d d r e s s B o o k dot C o n t a c t s A d d r e s
\item Mine is here backslash backslash g a b e h a l l hyphen m o t h r a backslash S v r underscore O f f i c e s v r
\item h t t p colon slash slash teams slash sites slash T A G slash default dot aspx As always , any feedback , comments ,
\item two thousand and five h t t p colon slash slash news dot com dot com slash i slash n e slash f d slash two zero zero three slash f d
\item backslash i n t e r n a l dot e x c h a n g e dot m a n a g e m e n t dot s y s t e m m a n a g e
\item I think Rich’s post highlights that we could have been more strategic about how the sum total of XBOX three hundred and sixtys were distributed .
\item 64X64 , 8K , one hundred and eighty four ASSEMBLY , DIGITAL VIDEO DISK DRIVE , INTERNAL , 8X ,
\item So we are back to Extended MAPI and C++ because . Extended MAPI does not have a dual interface VB or VB .Net can read .
\item Thanks , Borge Trongmo Hi gurus , Could you help us E2K ASP guys with the following issue ?
\item Thanks J RGR Are you using the LDDM driver for this system or the in the build XDDM driver ? 12
\item Btw , you might remember me from our discussion about OWA automation and OWA readiness day a year ago .
\item empidtool . exe creates HKEY\_CURRENT\_USER Software Microsoft Office Common QMPersNum in the registry , queries AD , and the populate the registry with MS employment ID if available else an error code is logged .
\item Thursday, via a joint press release and Microsoft AI Blog, we will announce Microsoft’s continued partnership with Shell leveraging cloud, AI, and collaboration technology to drive industry innovation and transformation.
\item Actress Fan Bingbing attends the screening of ’Ash Is Purest White (Jiang Hu Er Nv)’ during the 71st annual Cannes Film Festival

\end{enumerate}

\end{document}